\documentclass[aps,pro,twocolumn,showpacs,floatfix]{revtex4-1}

\usepackage[]{graphicx}
\usepackage{epstopdf}
\usepackage{amsmath,amssymb}
\usepackage{upgreek}
\usepackage[utf8]{inputenc}
\usepackage[colorlinks, linkcolor=blue, citecolor=blue, urlcolor=blue,breaklinks=true]{hyperref}
\usepackage{soul}

\providecommand{\ket}[1]{\ensuremath{\left\vert #1\right\rangle}}
\providecommand{\bra}[1]{\ensuremath{\left\langle #1\right\vert}}

\providecommand{\comm}[2]{\left[#1,#2\right]}

\begin{document}

\title{Observation of the photon-blockade breakdown phase transition}
\author{J.~M.~Fink${}^{1,2}$, A.~Dombi${}^3$,  A.~Vukics${}^3$, A.~Wallraff${}^1$, P.~Domokos${}^3$\\[3pt]
{\footnotesize \it ${}^1$ Department of Physics, ETH Z\"urich, CH-8093 Z\"urich, Switzerland}\\
{\footnotesize \it ${}^2$ Institute of Science and Technology Austria, 3400 Klosterneuburg, Austria}\\
{\footnotesize \it  ${}^3$ Wigner Research Centre for Physics, H-1525 Budapest, P.O. Box 49., Hungary}
}
\date{\today}

\begin{abstract}
Non-equilibrium phase transitions exist in damped-driven open quantum systems, when the continuous tuning of an external parameter leads to a transition between two robust steady states.
In second-order transitions this change is abrupt at a critical point, whereas in first-order transitions the two phases can co-exist in a critical hysteresis domain. Here we report the observation of a first-order dissipative quantum phase transition in a driven circuit quantum electrodynamics (QED) system. It takes place when the photon blockade of the driven cavity-atom system is broken by increasing the drive power. The observed experimental signature is a bimodal phase space distribution with varying weights controlled by the drive strength. Our measurements show an improved stabilization of the classical attractors up to the milli-second range when the size of the quantum system is increased from one to three artificial atoms. The formation of such robust pointer states could be used for new quantum measurement schemes or to investigate multi-photon quantum many-body phases.
\end{abstract}
\maketitle

The prototype of a nonlinear quantum system is the one described by the Jaynes--Cummings (JC) model of a two-level system coupled to a harmonic oscillator \cite{Jaynes1963}. This model corresponds to very high accuracy to cavity QED systems \cite{Haroche2006}, where atomic dipole transitions are coupled to quasi-resonant radiation modes of a resonator, or to circuit QED systems \cite{Wallraff2004}, where artificial atoms made of superconducting Josephson junctions are coupled to on-chip microwave resonators. The confinement of the photon into a small resonator volume results in a very strong coupling to the atom. Although the radiation mode spectrum is the well-known harmonic ladder, the coupling changes this to the significantly anharmonic JC spectrum, which allows for designing nonlinear processes within the low-intensity quantum domain and by means of only a single atom as a non-linear medium in the resonator. 

For monochromatic external driving with moderate power, the well-resolved resonances within the anharmonic JC spectrum \cite{Schuster2008Nonlinear,fink2008climbing,bishop2009nonlinear,shamailov2010multi} realize effectively a set of independent two-level systems each of which can be selectively addressed. When tuning the cavity driving to resonance with the lowest-lying excited state, at most  a single photon can be in the resonator. This effect was named photon-blockade  \cite{imamoglu1997strongly} in analogy with Coulomb blockade for electrons in a quantum well, and was experimentally demonstrated by observing photon anti-bunching in the transmitted radiation \cite{birnbaum2005photon,Faraon2008,Lang2011Observation}. Photon blockade is, in fact, more than the single-photon effect. The next manifold containing two energy quanta, can be used as a two-photon gateway \cite{Kubanek2008Twophoton}, and the concept can be further generalized to higher-order multi-photon transitions. 

The discrete quantum system with its  anharmonic spectrum cannot be excited away from the well-resolved resonances. It is then fully reflective and stays in a dim state close to the ground state. When increasing the power of the drive, the effective two-level system undergoes power broadening by saturation and the disjoint set of off-resonant frequency ranges shrinks until the broadened resonances merge into a single broad structure \cite{dombi2015bistability}. However,  even for very large drive powers, a photon-blockade frequency domain persists near the mode (here we assume, for simplicity, resonance between the cavity mode and the atomic transition). In this frequency range, interestingly, the further enhancement of the drive power breaks the photon blockade in a first order phase transition \cite{carmichael2015}. 

In first-order phase transitions, the two stable phases can co-exist in a critical range of parameter values. Quite typically, it is the direction from which the control parameter is tuned into the critical domain that determines which phase the system is prepared in. At a quantum mechanical level, this well-known hysteresis effect gives rise to continuously varying weights in a bimodal phase space distribution of the quantum system, as one of the classical attractors takes over by tuning the control parameter. This kind of transition takes place between the reflective vacuum state and a highly excited semiclassical state in the breakdown of the photon blockade regime \cite{dombi2015bistability,carmichael2015}. 

The coexistence of the two robust states with different macroscopic attributes is qualitatively different from a second-order phase transition which can also exist in non-equilibrium settings, i.e. in a damped-driven open quantum system. In second-order transitions the two phases are sharply separated by a critical point where the quantum fluctuations must diverge so that the two phases with different symmetries can be matched. The self-organization of a Bose-Einstein condensate in an optical cavity \cite{Nagy2008Selforganization} belongs to this class and has been experimentally observed \cite{Baumann2010Dicke}. 

A first-order dissipative quantum phase transition has been predicted in a wide variety of systems such as the vortex formation of a stirred Bose-Einstein condensate \cite{Dagnino2009Vortex}, or the deflection induced coupling between an electron spin and a mechanical resonator \cite{palyi2012spin}. Very recently, spatial symmetry breaking and hysteresis have been observed in the equilibrium state of an ultra-cold gas trapped in a double well potential \cite{Trenkwalder2016}, but the associated bimodal phase space distribution of a first-order dissipative phase transition has not yet been observed, to the best of our knowledge.

In the case of the driven JC model, highly nonlinear dynamics and multi-stability can occur in various parameter regimes. The phase transition effects are very distinct between the cases of a non-decaying or decaying two-level system, $\gamma=0$ or $\gamma\neq0$, since in the semiclassical limit the spin variable is confined to the surface of the Bloch sphere in the former case. 
For $\gamma\neq0$, the highly excited state involves a completely saturated two-level system, i.e. the semiclassical spin variable contracts into the origin of the Bloch sphere. This is the case of the absorptive optical bistability, which has been studied in cavity QED realizations of the JC model \cite{Rosenberger1991,Rempe1991Optical}. However, in an atomic cavity QED system the coupling strength does not allow for well resolved semiclassical attractors in the single-atom limit \cite{Savage1988,Kerckhoff2011Remnants}. 

In contrast, for $\gamma=0$, the so-called spontaneous dressed state polarization \cite{alsing1991spontaneous} takes place in the resonantly driven resonant JC system via a second-order transition above a threshold drive strength. Its remnants in the form of phase flipping have been observed in the homodyne measurement of light transmitted through a Fabry-Perot cavity, interacting with laser cooled $^{133}$Cs atoms \cite{Armen2009}. Away from resonant driving, a first order transition via bistability is expected \cite{carmichael2015}. For $\gamma=0$, this bistability extends to the dispersive limit of the JC model, which has been proposed for high fidelity qubit readout in the context of circuit QED \cite{Ginossar2010}. In this dispersive limit the JC nonlinearity has been shown to improve single-qubit measurements \cite{Reed2010High,bishop2010response, Boissonneault2010Improved}, similar to latching readouts with nonlinear cavities based on additional Josephson circuit elements \cite{Murch2012Quantum}.  

In microscopic quantum systems, such as a single-atom cavity QED system, one can refer to phases because one of the subsystems is a harmonic oscillator with infinitely many states. Therefore, the highly-excited part of the JC spectrum is quasi-harmonic and can host coherent states, or weakly phase-stretched coherent states due to the residual anharmonicity. These states can be robust stationary states, i.e. semiclassical attractors of the damped-driven system. The linearity of the high-lying states vs. the nonlinearity of the low-lying states of the JC model forms the basis for a quantum phase transition. A non-stationary, dynamical phase transition has been observed in a coupled JC dimer \cite{raftery2014observation}, but the detailed study of the stationary steady-state transition is hindered both experimentally and numerically by the very large critical photon numbers in resonantly driven JC model. This obstacle can be avoided with a relatively simple trick. We show that coupling a third atomic level into the dynamics drastically changes the excitation route to the highly excited attractor and the transition regime between the two phases can be explored in detail.

\begin{figure}[t]
\begin{center}
\includegraphics[width=1\columnwidth]{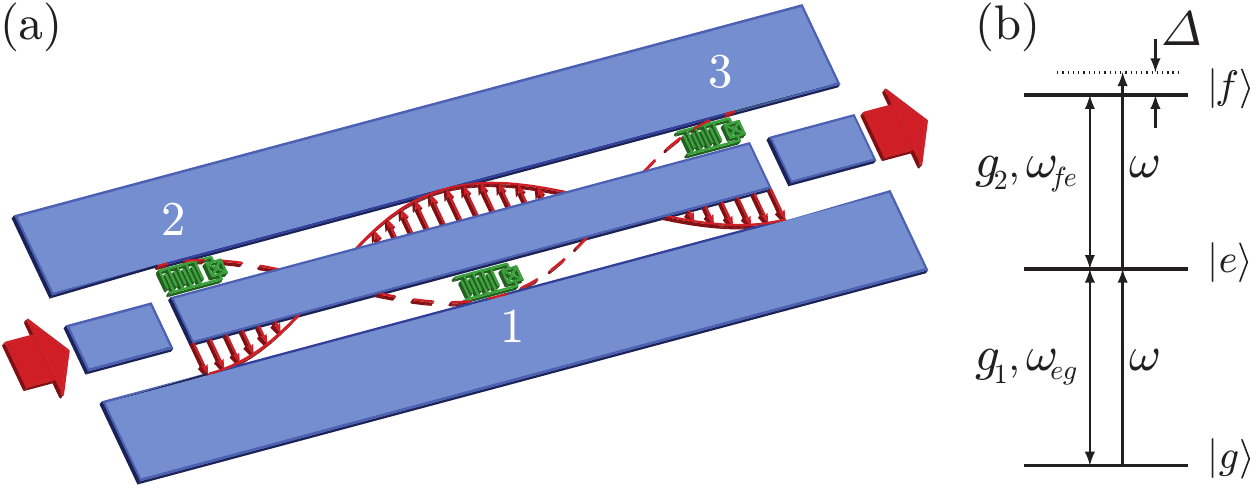}
\caption{(a) Scheme of the driven coplanar waveguide resonator (blue) coupled to up to 3 transmon artificial atoms (green). An external drive (red) is applied via the input capacitor, and the coherent transmission is detected via the output capacitor. (b) Level scheme of one artificial atom indicating the parameters of the Hamiltonian in Eq.~(\ref{eq:Hamiltonian}).}
\label{fig:scheme}
 \end{center}
\end{figure}

A suitable experimental setting to realize the three-level system and the necessary strong coupling strength is superconducting circuit QED, which offers large dipole coupling strengths and long coherence times of multiple superconducting artificial atoms embedded in a high-quality on-chip microwave cavity \cite{Fink2009Dressed}. Our experimental device and setup is similar to the one in Ref.~\cite{Mlynek2012} and consists of three frequency tunable superconducting artificial atoms positioned at the anti-nodes of the first harmonic voltage standing wave resonance of a coplanar superconducting resonator at frequency $\omega_\mathrm{c}/(2\pi)=7.024~\mathrm{GHz}$, see Fig.~\ref{fig:scheme}(a). 

We use capacitively shunted superconducting charge qubits, or transmon qubits \cite{Koch2007Transmon}, which have a flat charge dispersion but also a limited absolute anharmonicity of $\omega_\mathrm{fe}-\omega_\mathrm{eg}\simeq -E_\mathrm{C}/\hbar$. Here, $|g\rangle$, $|e\rangle$ and $|f\rangle$ are the first three levels of the artificial atom, see Fig.~\ref{fig:scheme}(b) which all take part in the dynamics of the presented experiments. $E_\mathrm{C}/h=(459, 359, 358)~\mathrm{MHz}$ are the charging energies and $g_1/(2\pi)=(-52.7, 55.4, 55.8)~\mathrm{MHz}$ the single photon $|g\rangle$ to $|e\rangle$ transition to resonator coupling strengths of the atoms 1, 2 and 3. Similar to a harmonic oscillator, the coupling strength of the second excited level can be approximated \cite{Koch2007Transmon} as $g_2\simeq\sqrt{2}g_1$. Using externally applied and locally concentrated magnetic fields we can control the flux $\Phi$ through the individual transmon SQUID loops and change their Josephson energy $E_\mathrm{J}(\Phi)$. This allows to independently control the transmons' transition energies $\hbar \omega_\mathrm{eg}\simeq\sqrt{8 E_\mathrm{C}E_\mathrm{J}(\Phi)}-E_\mathrm{C}$ and in particular to tune them in and out of resonance with the microwave resonator with a full width half maximum linewidth of $2\kappa/(2\pi)=0.47$~MHz. For the most part of this manuscript we study the situation where two of the three artificial atoms are far detuned ($\omega_\mathrm{eg} \ll \omega_\mathrm{r}$) and not interacting with the microwave resonator.

The Hamiltonian of the system in a frame rotating at the angular frequency $\omega$ of an external drive reads
\begin{multline}
\label{eq:Hamiltonian}
H/\hbar =  \Delta_1 \ket{g}\bra{g} - \Delta_2 \ket{f}\bra{f}  - \Delta_\mathrm{c} a^\dagger a + \eta (a + a^\dagger) \\
+ g_1 (a^\dagger\ket{g}\bra{e}+\ket{e}\bra{g}a) + g_2 (a^\dagger\ket{e}\bra{f}+\ket{f}\bra{e}a) \,,
\end{multline}
where the detunings are defined as $\Delta_1 = \omega - \omega_\mathrm{eg}$, $\Delta_2 = \omega - \omega_\mathrm{fe}$, and $\Delta_\mathrm{c} = \omega - \omega_\mathrm{c}$, and $\eta$ is the effective amplitude of the drive field.  For the drive frequency, we will consider the special case $\Delta_1=0$, as shown schematically in Fig.~\ref{fig:scheme}(b). We assume resonance also between the resonator mode and the $|g\rangle\leftrightarrow|e\rangle$ transition, i.e., $\Delta_\mathrm{c} =0$. The remaining detuning is denoted by $\Delta_2 \equiv \Delta$, which is then simply the charging energy $E_\mathrm{C}/\hbar$. The dissipative processes are accounted for by the following Lindblad terms in the  Master equation of the density matrix 
\begin{multline}
\label{eq:Master}
\dot{\rho} = -\tfrac{i}{\hbar} \comm{H}{\rho} + \kappa\left( 2 a \rho a^\dagger - a^\dagger a \rho - \rho a^\dagger a \right) \\ + \gamma_{\parallel} \left( 2 \ket{g}\bra{e} \rho \ket{e}\bra{g} - \ket{e}\bra{e} \rho - \rho \ket{e}\bra{e}  \right) \\ + \gamma_{\parallel} \left( 2 \ket{e}\bra{f} \rho \ket{f}\bra{e} - \ket{f}\bra{f} \rho - \rho \ket{f}\bra{f}  \right) \\  + 8 \gamma_{\perp}\left( \left[ \sum_{i=g,e,f} \ket{i}\bra{i} \rho \ket{i}\bra{i} \right]  - \rho\right)\,.\end{multline}
The rate $2 \kappa$ characterizes the cavity-photon loss due mainly to the photon out-coupling to propagating modes. There can be  population and polarization damping in the three-level system, which have the rates $\gamma_{\parallel}$ and $\gamma_{\perp}$, respectively. In the theoretical description we assume zero-temperature reservoirs.

Obviously, without the third excited state $\ket{f}$ (formally, we can set  $g_2=0$) the model describes the damped-driven JC model. The vacuum Rabi splitting $2g_1$ exceeds by $\sim2$ orders of magnitude the linewidth of the dressed state resonances,  $g_1\gg\kappa, \gamma_\perp, \gamma_\parallel$.  Even the multi-photon transitions to higher-lying dressed states are far out of resonance when the system is driven at the bare resonator  frequency, $\Delta_1=\Delta_\mathrm{c}=0$. This setting favors the photon-blockade dim state up to large drive strengths. 

\begin{figure}[t]
 \begin{center}
 \includegraphics[width=1.0\columnwidth]{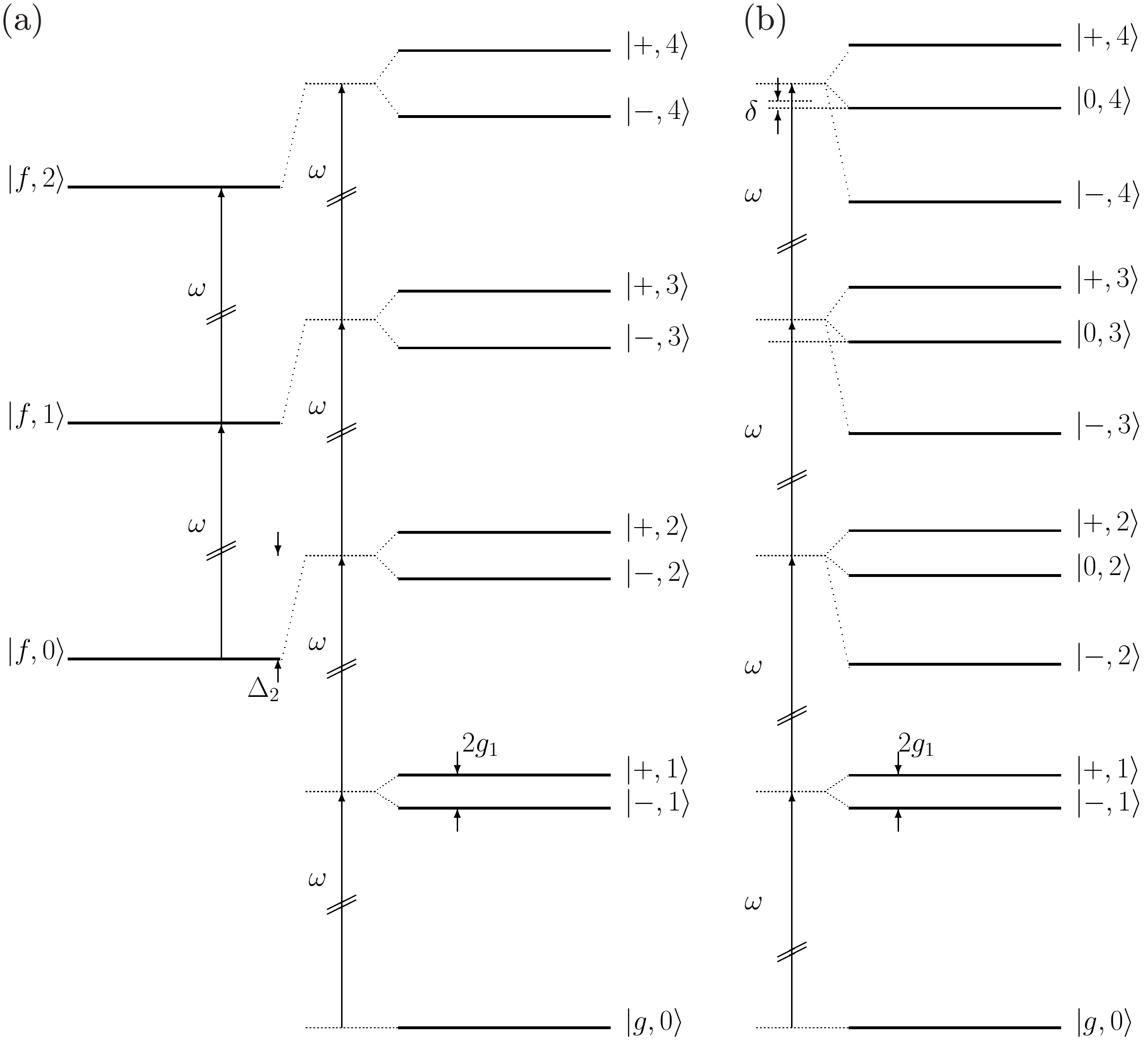} 
\caption{(a) The spectrum of the coupled atom-cavity system if the third  atomic state $|f\rangle$ is only weakly coupled to the middle one $|e\rangle$, that is, $g_2\ll g_1$. In (a) the $\ket{\pm, n}$ denote the usual two-level atom JC model dressed states which combine the states $\ket{g,n+1}$ and $\ket{e,n}$. (b) The spectrum of the coupled atom-cavity system in the experiment is $\Delta_1 = 0, \Delta_2 \simeq E_\mathrm{C}/\hbar$ and $g_2 \simeq \sqrt{2} g_{1}$. In (b) the multiplets $\ket{\{-,0,+\},n}$ correspond to the eigenstates of Eq.~\ref{eq:Hamiltonian} in the manifold of a number of $n$ excitations. $\delta$ represents the mismatch of $\omega$ with respect to the transition frequency, which is an order of magnitude larger than the linewidth.}
\label{fig:spectra}
 \end{center}
\end{figure}

Assume now that there is a very weak coupling of the state $\ket{e}$ to $\ket{f}$, that is $0 \neq g_2 \ll g_1$. Because of this latter relation, the JC spectrum associated with the two-level system $\ket{g}$ and $\ket{e}$ is only slightly perturbed and the corresponding dressed states $\ket{\pm, n}$ can be easily identified in the full spectrum, as shown in Fig.~\ref{fig:spectra}(a).  The ground state is the $\ket{g,0}$ state and the first excited state manifold is $\ket{\pm,1}$ exhibiting the vacuum Rabi splitting $2g_1$. Starting from the second excited manifold, there is a third level in each manifold which weakly couples to the JC dressed states. In these levels, because of the small coupling  $g_2\ll g_1$ the dominant component is the atomic excitation $\ket{f}$. Therefore, the level spacing between these third levels $\ket{f,n}$ and the adjacent ones $\ket{f,n\pm1}$,  is uniform and resonant with the photon frequency $\omega$.  This equidistant ladder in the spectrum has dramatic consequences. 

When the cavity is resonantly driven, the off-resonant states $\ket{\pm,2}$ are weakly populated by two-photon transitions. A small fraction of this anyway small population leaks into this harmonic part of the spectrum, which is then resonantly driven within the  manifold $\ket{f,n}$. This component of the wavefunction is thus coherently displaced in the photon mode on the time scale of $\kappa^{-1}$. The displacement is not interrupted by atomic decay since the polarisation damping keeps the atomic state $\ket{f}$ intact, and the population decay from $\ket{f}$ to $\ket{e}$ is negligible on the excitation time scale, $\gamma_\parallel^{-1} \gg \kappa^{-1}$. The displacement is counteracted by cavity loss and leads to a steady state, which is a coherent state with an amplitude $\eta/\kappa$ and phase locked to the phase of the driving field. None of the loss processes leads out of this coherent excited state: (i) the cavity loss, together with the driving strength, sets the amplitude and does not break coherence; (ii) once the  photon number is large enough, the change of the atomic state does not matter because the Jaynes--Cummings spectrum approaches an harmonic one also in the manifolds of states $\ket{+,n}$ and $\ket{-,n}$, and the phase is locked again to the phase of the drive $\eta$. 

Just like in optical pumping, the total state of the system is gradually pumped into this \emph{trapping state}, a coherent state of the resonator mode combined with a full mixture of the atomic states. The bottleneck formed by the initial two-photon transition and the weak $\ket{\pm,2} \rightarrow \ket{f,0}$ coupling increases only the time the system is pumped into this trapping state. Rather surprisingly, the three-level atom in a cavity with the third level weakly coupled to the other two leads thus to a steady-state transmission spectrum identical to that of an empty cavity. 

When the coupling $g_2$ is increased to $g_1 \lesssim g_2$ the atomic state $\ket{f}$ couples significantly to the other states $\ket{g}$, $\ket{e}$ to form an anharmonic spectrum shown in Fig.~\ref{fig:spectra}(b). The resonant driving $\Delta=0$ becomes off-resonant with respect to the anharmonic spectrum.  For the transitions among the middle states in the triplets, the multi-photon excitation is hindered by the polarisation damping which mixes the dressed-states  within the triplet on the same time scale as the cavity loss $\gamma_\perp \sim \kappa$.

\begin{figure}[t]
 \begin{center}
\includegraphics[width=1\columnwidth]{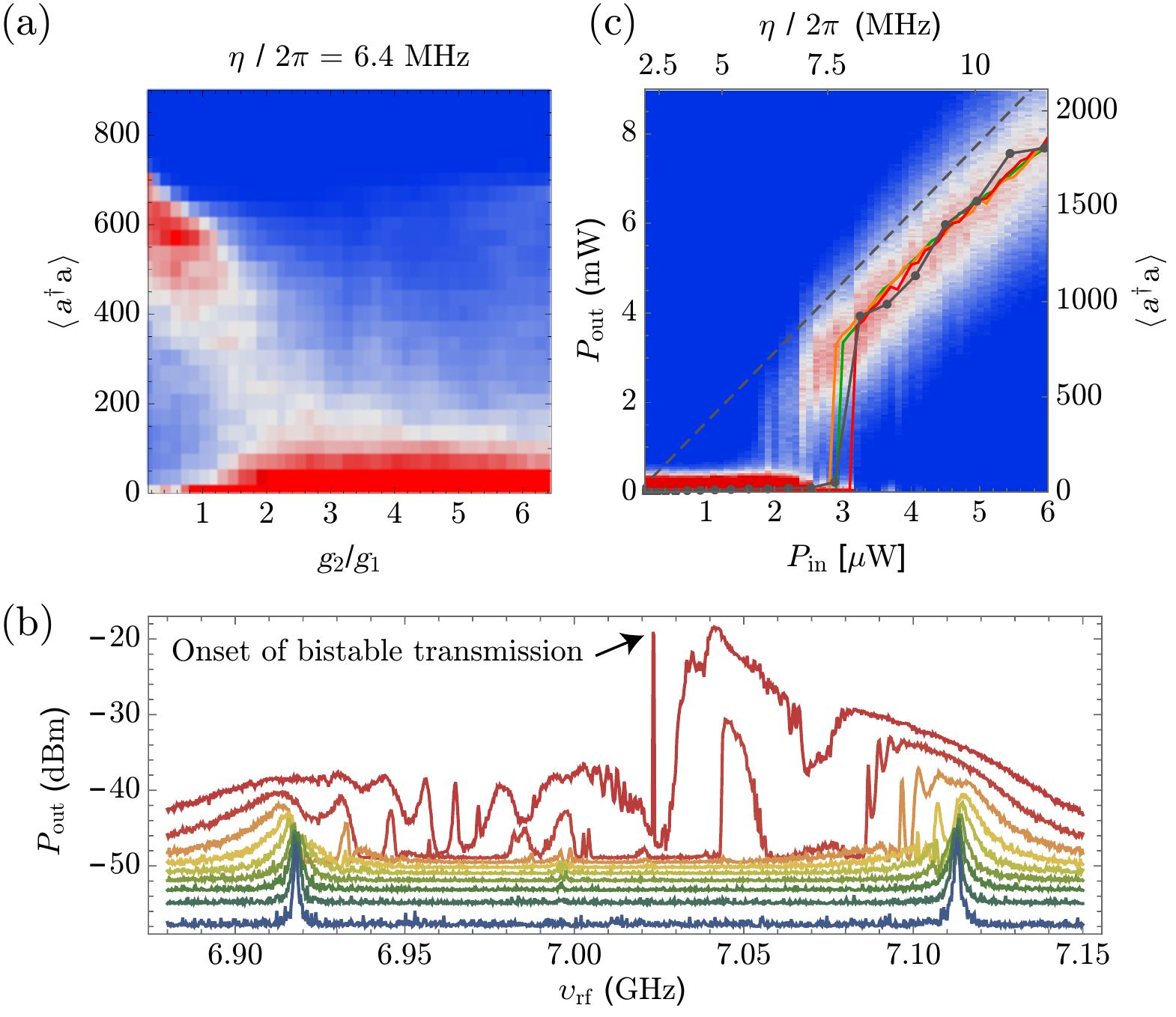}
\caption{
(a) Simulated histogram of the output intensity as a function of the coupling constant $g_2$ between the two excited atomic states $\ket{e}$ and $\ket{f}$ for $\eta /(2\pi)= 6.4$~MHz with red representing high probability and blue indicating zero probablility. 
(b) Measured vacuum Rabi spectra for different input powers $P_\mathrm{in}=1.8\times10^{-4}\mu\mathrm{W}$ (blue) to $1.8\mu\mathrm{W}$ (red) with all 3 atoms in resonance with the cavity. The spectra are offset by $1.6$~nW for better visibility. $P_\mathrm{in}$ and $P_\mathrm{out}$ refer to the microwave generator output and digitizer input power respectively.
(c) Measured histogram of the detected power as a function of the cavity input power for a single transmon (density plot). The most likely photon numbers (line plots) are extracted from this measurement (red) and two similar measurements taken with 2 (orange) and 3 qubits (green) in resonance with the cavity mode. Simulation results for the single qubit case are shown with connected black symbols for comparison. The dashed line is for reference and represents the response of the empty cavity with $\langle a^\dagger a\rangle=(\eta/\kappa)^2$.}
\label{fig:explore_g}
\end{center}
\end{figure}

Figure \ref{fig:explore_g}(a) presents a histogram of the calculated transmitted field intensity as a function of $g_2$ in the range between the two limiting cases, $g_2=0$ and $g_1 \lesssim g_2$. The histogram is prepared from the ensemble of intensity values recorded at many randomly chosen instants while the system is in steady state. The numerical simulations were performed using the C++QED framework \cite{VukicsCppQEDa,VukicsCppQEDb}, which solves the master equation  (\ref{eq:Master}) by unravelling it into a set of Monte Carlo wave-function trajectories. 
As expected, small $g_2$ gives rise to a transmission of an resonantly driven empty cavity, the photon number fluctuates around the mean $\langle a^\dagger a \rangle \approx \left(\eta/\kappa\right)^2 = 700$. Larger fluctuations and the residual population in the low photon states, close to the origin $g_2\approx 0$ of the plot,  is merely a finite simulation time effect.  On the other hand, it is also confirmed that the three-level atom with $g_2 > g_1$  switches off the transmission. It must be larger than $g_2 \sim g_1$, because the bare level $\ket{f}$ is significantly detuned from the resonance with the drive.

The histogram in Fig.~\ref{fig:explore_g}(a) reflects a transition domain manifested by a bimodal distribution of intensities. Experimentally we are exactly in this transition regime with $g_2/g_1\simeq\sqrt{2}$. Figure \ref{fig:explore_g}(b) presents measured vacuum Rabi spectra for different input probe powers varied over 4 orders of magnitude where all 3 transmons are in resonance with the cavity mode (all indicated power values refer to generated / detected powers at room temperature). At low input powers corresponding to much less than a single intra-cavity photon on average, we observe the well known splitting of the coupled multi-qubit single-photon state \cite{Agarwal1984}. Here we observe no additional resonances which validate that the system is initialized in its quantum ground state rather than a thermal state \cite{Fink2009b}. At intermediate powers we observe a rich structure in the transmission spectrum, which is determined by multi-qubit multi-photon transitions. The power broadening of the multi-photon resonances can be observed. The situation is even more complicated due to the additional transmon levels which lead to an asymmetry of the observed spectra \cite{fink2008climbing,bishop2009nonlinear}. More importantly however, the frequency region around the bare cavity frequency remains dim over a large range of input powers. At a certain power, corresponding to about 500 intra-cavity photons, we observe a sudden narrow transmission peak at the bare cavity frequency. This transmission peak is observed to be switching for a certain range of constant input powers. At even higher input powers the spectrum (not shown) resembles that of an empty cavity without atoms \cite{Fink2010,Reed2010High}.

\begin{figure*}[t]
 \begin{center}
\includegraphics[width=2\columnwidth]{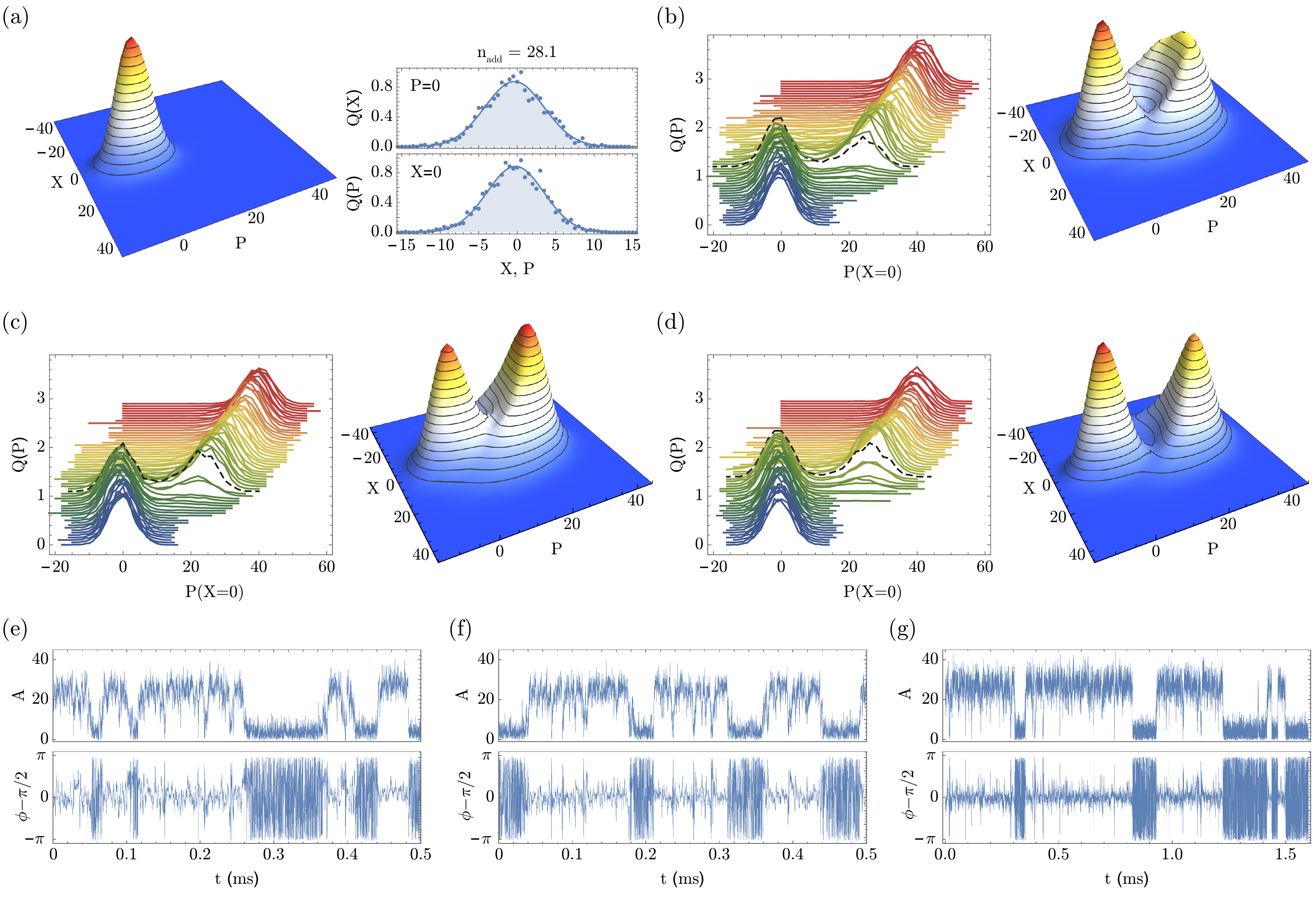}
\caption{
(a) Measured Q-function of the vacuum field for very low drive power $\eta/(2\pi)=1.4$~MHz in the photon blockade regime (left) and Gaussian fit along the two field quadratures X and P. The fit reveals a thermal state due to the added amplifier noise with $n_\mathrm{add}=28.1$. 
(b) The waterfall plot shows line cuts of the measured Q-functions where $X=0$ as the drive strength is increased from $\eta/(2\pi)=1.4$ to 11~MHz for a single transmon in resonance with the cavity (left). A vertical offset of 0.05 between the line plots has been used for better visibility. The complete Q function for the critical drive strength $\eta/(2\pi)=7.1$~MHz, which is indicated by the dashed black line on the left. 
(c) The same measurement for transmons 1 and 2 in resonance and the critical drive strength $\eta/(2\pi)=6.8$~MHz.
(d) The same for transmons 1, 2 and 3 in resonance and the critical drive strength $\eta/(2\pi)=7.7$~MHz.
(e)-(g) The real-time single-shot record of the transmitted output field amplitude and phase for one (e), two (f) and three (g) atoms in resonance with the cavity.}
\label{fig:Qfunction}
\end{center}
\end{figure*}

We record the cavity transmission as a function of input power at the bare cavity frequency for the case of one, two and three transmons in resonance with the cavity mode. In each case the transmitted tone is amplified with a commercial HEMT amplifier and digitized with a time-resolution of 40~ns for 1.6~ms. This corresponds to a filter bandwidth of 25 MHz, much smaller than the vacuum Rabi splitting and small enough to filter out most low lying multi-photon transitions. The density plot shown in Fig.~\ref{fig:explore_g}(c) shows the likelihood to detect a particular output power as a function of a large range of input powers, both referenced to the generator / digitizer outside of the dilution refrigerator. 

The measured histogram clearly shows an input power region between 2 and 3 $\mu$W with two distinct but equally likely solutions. For smaller powers we observe no transmission (photon blockade) and for higher powers the cavity transmission scales linearly with the input power (empty cavity solution). The red line-plot indicates the most likely output power for this measurement and agrees very well to the simulated results shown in black. The simulation results are based on independently measured sample parameters, see also Ref.~\cite{Mlynek2012}, and provide a calibration for the drive strength $\eta$ shown on the top horizontal axis of Fig.~\ref{fig:explore_g}(c) and the mean intra-cavity photon number $\langle a^\dagger a\rangle$ shown on the right hand vertical axis. This calibration agrees well with the estimated total attenuation (86~dB) of the input line. The orange and green lines show the results of the identical measurements with two and three atoms in resonance with the cavity mode. The characteristic switching power is slightly shifted, but no other differences are apparent between the single and multi-atom case.

A full characterization of the steady-state of the resonator field can be given by the Husimi Q quasiprobability distribution, see Fig.~\ref{fig:Qfunction}. For small drive strength the system is in the photon blockade regime and the Q function is that of the vacuum state. Due to the added noise of the amplifier chain we expect a convolution of the Q-function vacuum state with a thermal state and a Gaussian fit to the measured distribution (see Fig.~\ref{fig:Qfunction} a) yields an added noise photon number of 28.1 as referenced to the cavity output. This corresponds to a total system noise temperature of $9.5$~K, in line with our expectations due to losses in the output line and the used amplifiers' noise figures. 

Figure~\ref{fig:Qfunction} (b), (c) and (d) show the experimental results for one, two and three atoms in resonance, respectively. The waterfall plots explicitly show the photon blockade breaking transition indicated by the sudden change of the Q function peak position as the drive power is increased from low (blue) to high drive strength (red). The right hand plots depict the entire Q function close to the critical point, i.e. for the data sets marked by dashed black lines on their left. In all three cases the Q function shows a clear bimodal structure with two solutions. Due to the large intra-cavity photon number in the bimodal switching regime of $\sim 700$, we can clearly resolve the double-peaked Q-function demonstrating the mixture of the vacuum state and the highly excited state with well-defined amplitude and phase. From the better resolved Q function in the 3 atom case compared to the single and two atom cases, it appears that that the phase stability is improved for larger systems.

The observed (bi)stability can be maintained only for a finite time and there is a substantial difference with respect to the classical bistability. Within the quantum domain, such as in the microscopic medium composed by a single atom, the branches are not stable on macroscopic time scales. This is shown in Fig.~\ref{fig:Qfunction} (e-g) which presents the time evolution of the system in steady-state. The system flips between two quasi-stationary solutions corresponding to the two components of the bimodal Q function: the vacuum with an undefined phase and an excited coherent state. The well resolved plateaus reveal that the dwell time in each of these components is longer then the microscopic time scales characteristic for equilibration. The typical time scale is expected to depend on the system size and would increase to infinity for a macroscopic system. Experimentally we find a much longer characteristic switching time of $\sim 0.5$~ms when all 3 transmons are tuned in resonance with the microwave resonator, see Fig.~\ref{fig:Qfunction} (g), compared to the time dynamics with only one (panel (e)) or two atoms (panel (f)), where the typical switching time is on the order of $\sim 0.1$~ms. In all cases the typical time scales exceed by far the cavity and qubit coherence times of 0.1 to 0.4~$\mu$s \cite{Mlynek2012}. 

In summary, we have studied the non-equilibrium first-order phase transition that takes place when a strongly driven quantum systems breaks through the non-resonant, reflective phase into a transmissive excited one. We have shown both theoretically and experimentally that the presence of an additional near resonant atomic level dramatically changes the system dynamics and allows us to access the excited phase. In a given range of driving strengths, the transition gives rise to a mixture of the two phases, and the quantum system is switching between the two corresponding states. The increase of the dwell time as the size of the system is enlarged, from one to three atoms in our experiment, has been observed, and points to the stabilization of the corresponding phases in the classical limit.

Our results represent an important first step towards exploring many body physics in the many photon - high drive power regime. A detailed understanding of quantum phase transitions in circuit QED \cite{Feng2015, Tian2010} could play an important role for the controlled simulation and stabilization of many body quantum states. In the future, one could study entanglement \cite{Rogers2015} and the details of the time dynamics using auto-correlation measurements \cite{Lang2011Observation}. With better qubit coherence times and quantum limited amplifiers, phase multi-stability \cite{Delanty2011} and squeezing \cite{Peano2010Dynamical} may be realizable. Another potential direction is the application as an ultra-low energy classical set reset flip-flop memory for ultra low power signal processing in photonics \cite{Kerckhoff2011thedressedatom} or classical front end processing for quantum computers \cite{Andersen2015}. 

We thank A. Blais for comments on the manuscript. This work was supported by ETH Zurich, IST Austria, the Hungarian Academy of Sciences (Lend\"ulet Program, LP2011-016) and the National Research, Development and Innovation Office (K115624). A.V. acknowledges support from the János Bolyai Research Scholarship of the Hungarian Academy of Sciences.

\bibliography{references}
\end{document}